\documentclass[english,letterpaper,twocolumn,showpacs,pra,aps]{revtex4}
\usepackage{amssymb}
\usepackage{amsmath}%
\usepackage{verbatim}
\usepackage{amsfonts}%
\usepackage{graphicx}
\usepackage{epstopdf} 

\newcommand{\be}[0]{\begin{equation}}
\newcommand{\ee}[0]{\end{equation}}
\newcommand{\ba}[0]{\begin{align}}
\newcommand{\ea}[0]{\end{align}}
\newcommand{\ban}[0]{\begin{align*}}
\newcommand{\ean}[0]{\end{align*}}

\makeatletter

\large  

\input epsf

\makeatother

\usepackage{babel}
\makeatother
\begin{document}

\title{Electromagnetic properties of thin metallic films}

\author{Luke S. Langsjoen, Amrit Poudel, Maxim G. Vavilov, and Robert Joynt}

\affiliation{Department of Physics, University of Wisconsin, Madison, Wisconsin 53706, USA}

\begin{abstract}
We compute the electromagnetic fluctuations due to evanescent-wave Johnson noise in the vicinity of a thin conducting film, such as a metallic gate or a 2-dimensional electron gas.  This noise can decohere a nearby qubit and it is also responsible for Casimir forces.  We have improved on previous calculations by including the nonlocal dielectric response of the film, which is an important correction at short distances.  Remarkably, the fluctuations responsible for decoherence of charge qubits from a thin film are greatly enhanced over the case of a conducting half space.  The decoherence times can be reduced by over an order of magnitude by decreasing the film thickness.  This appears to be due to the leakage into the vacuum of modes that are well localized in the perpendicular direction.  There is no corresponding effect for spin qubits (magnetic field fluctuations).  We also show that a nonlocal dielectric function naturally removes the divergence in the Casimir force at vanishing separation between two metallic sheets or halfspaces.  In the separation regime where a local and nonlocal treatment are noticeably distinct, the Casimir attraction between two thin sheets and two halfspaces are practically indistinguishable for any physical film thickness.
\end{abstract}

\maketitle
\section{Introduction}
Thin metallic films are being used in an increasing number of nano-technological applications.  Semiconductor qubit architectures use conducting gates to isolate, manipulate and read the qubit.  While these conducting device elements are essential to the functionality of the qubit, they also give rise to an inevitable source of decoherence through evanescent-wave Johnson noise \cite{henkel}.  The top gates in accumulation-mode qubit architectures in particular are well approximated by a thin metallic film, and an accurate calculation of the decoherence times in these devices will require a detailed treatment of the electromagnetic properties of the films \cite{borselli}.  Thin films can also supply a desired or undesired source of heat transfer in micro-mechanical devices.  Free-standing conducting films will also experience stiction forces from nearby device elements through the Casimir effect. 

The magnitude of heat transfer, the Casimir effect, or the qubit decoherence rate can all be obtained once the reflection coefficients of the film, $r_p$ and $r_s$, have been calculated. $r_p$ is the reflection coefficient for incident light with an electric field that is polarized in the plane of incidence, while $r_s$ has its electric field polarized perpendicular to the incident plane. In this paper we present a detailed derivation of these coefficients for the case of a general nonlocal dielectric function, and use them to obtain quantitative calculations of the effects mentioned above.  The thin film reflection coefficients are found to exhibit important differences from those from a halfspace. These same nonlocal reflection coefficients have been found previously by Esquivel-Sirvent and Svetovoy \cite{esquivel} in the context of the Casimir effect. The main contribution of the present work is to supply a different derivation of the result, and to apply the coefficients to a broader variety of problems.  Of central importance is our result that a nonlocal treatment removes all divergences in the size of these effects as the surface of the film is approached.

This paper is organized as follows.  Section \ref{derivation} derives our expression for the reflection coefficients of a thin film in a nonlocal treatment. Section \ref{app} then applies these coefficients to give a quantitative analysis of qubit decoherence from evanescent-wave Johnson noise, heat transfer, and the Casimir force.  Section \ref{limit} shows how our reflection coefficients reduce to the cases of a local response and a conducting halfspace when the appropriate limits are taken.  Finally, in Section \ref{conc} we present our conclusions.

\section{Derivation}
\label{derivation}
We consider an infinite metallic sheet with nonlocal dielectric response whose surfaces are located at $z=-a$ and $z=0$. To derive the reflection coefficients, we generalize the treatment by Ford and Weber in \cite{fordweber} of a halfspace in the semiclassical infinite barrier model.  The fields inside the sheet satisfy Maxwell's equations:

\begin{align}
\nabla \cdot\textbf B(\textbf{r})=0~,~~\nabla\times\textbf B(\textbf{r})+i\frac{\omega}{c}\textbf D(\textbf{r})=\frac{4\pi}{c}\textbf{j}(\textbf{r}),\nonumber\\
\nabla\cdot\textbf D(\textbf{r})=4\pi\rho(\textbf{r})~,~~\nabla\times\textbf E(\textbf{r})-i\frac{\omega}{c}\textbf{B}(\textbf{r})=0~,
\end{align}
where we consider fields varying harmonically in time at frequency $\omega$, $\textbf{E}(\textbf{r},t)=\textbf{E}(\textbf{r})e^{-i\omega t}$. If we define the Fourier modes of all field quantities as $\textbf{E}(\textbf{r})=\int d\textbf{k}\textbf{E}(\textbf{k})\exp(i\textbf k\cdot \textbf r)/(2\pi)^3$, etc., a general nonlocal dielectric function will relate $\textbf D$ and $\textbf E$ by
\be
\textbf D(\textbf k)=\epsilon_l(k,\omega)\hat{\textbf{k}}\cdot\textbf E(\textbf k)\hat{\textbf{k}}+\epsilon_t(k,\omega)\left(\textbf E(\textbf k)-\hat{\textbf{k}}\cdot \textbf E(\textbf k) \hat{\textbf{k}}\right),
\ee
where we have separated the dielectric function into its longitudinal, $\epsilon_l$, and transverse, $\epsilon_t$, components. The reflection coefficients may be found through the surface impedances \cite{LLECM}, defined as
\begin{align}
Z^P(p,\omega)&=-\frac{4\pi}{c}\left\{\frac{\hat{\textbf{p}}\cdot\textbf{E}}{\hat{\textbf{z}}\times\hat{\textbf{p}}\cdot\textbf{B}}\right\}_{inside}\nonumber\\
Z^S(p,\omega)&=\frac{4\pi}{c}\left\{\frac{\hat{\textbf{z}}\times\hat{\textbf{p}}\cdot\textbf{E}}{\hat{\textbf{p}}\cdot\hat{\textbf{B}}}\right\}_{inside},
\end{align}
where the fields are evaluated at the inner surface of the metal. The reflection coefficients may then be written as
\be
r_p=\frac{4\pi q_1/\omega-Z^P}{4\pi q_1/\omega+Z^P}~,~~r_s=\frac{Z^S-4\pi\omega/c^2q_1}{Z^S+4\pi\omega/c^2q_1}.
\ee
 In the semiclassical infinite barrier model, it is assumed that the conduction electrons exhibit specular reflection at the boundary. In this model, the behavior of the fields inside a conducting halfspace are indistinguishable from the fields inside an infinite conductor with a current sheet at the location of the surface,
 \be
\textbf{j}(\textbf{r},t)=\textbf{J}\delta(z)e^{i(\textbf{p}\cdot\boldsymbol{\rho} -\omega t)} , ~~\hat{z}\cdot\textbf{J}=0 ,
\ee
where $\boldsymbol \rho$ is the position vector in the plane of the boundary, not to be confused with the electron density. For our case of a thin conducting film, the single current sheet is replaced by an infinite series of image current sheets:
\begin{align}
\textbf{j}(\textbf{r},t)&=\sum_{n=-\infty}^\infty\left(\textbf{J}_1\delta(z-2an)+\textbf{J}_2\delta(z-a(2n+1))\right)e^{i(\textbf{p}\cdot\boldsymbol{\rho}-\omega t)}\nonumber \\
\hat{z}\cdot\textbf{J}_1&=0 , ~~ \hat{z}\cdot\textbf{J}_2=0\nonumber 
\end{align}
where $\textbf{J}_1$ and $\textbf{J}_2$, which correspond to images of the right and left surface current sheets, respectively, must be of equal magnitude and either parallel or antiparallel.  Plugging this current source into Maxwell's equations allows us to solve for the electric and magnetic fields inside the metal
\begin{align}
\textbf{E}(z)=&\frac{2\pi}{i\omega a}\sum_{n=-\infty}^\infty\Bigg(\left(\frac{\textbf{J}_1-(\textbf k\cdot\textbf{J}_1)\textbf k/k^2}{\epsilon_t-c^2k^2/\omega^2}+\frac{(\textbf{k}\cdot\textbf{J}_1)\textbf k}{k^2\epsilon_l}\right)\nonumber\\
&+(-1)^n\left(\frac{\textbf{J}_2-(\textbf k\cdot\textbf{J}_2)\textbf k/k^2}{\epsilon_t-c^2k^2/\omega^2}+\frac{(\textbf{k}\cdot\textbf{J}_2)\textbf k}{k^2\epsilon_l}\right)\Bigg)e^{iqz}\nonumber\\
\textbf{B}(z)=&\frac{2\pi c}{i\omega^2a}\sum_{n=-\infty}^\infty\left(\frac{\textbf{k}\times\textbf{J}_1}{\epsilon_t-c^2k^2/\omega^2}+\frac{\left(\textbf{k}\times\textbf{J}_2\right)(-1)^n}{\epsilon_t-c^2k^2/\omega^2}\right)e^{iqz}
\end{align}
It can be seen by inspection that $\textbf{J}_1=\textbf{J}_2$ corresponds to field components whose wavelength in the $z$-direction is an integer fraction of the thickness $a$, while $\textbf{J}_1=-\textbf{J}_2$ corresponds to wavelengths in the $z$-direction that are half-integer fractions of $a$. Comparison to Ford and Weber shows that the fields within a thin film differ from those within a halfspace by replacing the integral over a continuous $q$ by a summation over a discrete $q_n=2n\pi/a$ or $q_n=(2n+1)\pi/a$ depending on whether $\textbf{J}_1=\textbf{J}_2$ or $\textbf{J}_1=-\textbf{J}_2$, respectively. The reflection coefficients are then obtained by summing the contribution from both cases. 
\begin{align}
\label{rpnla}
r_{p}=&\frac{1}{2}\displaystyle\sum_{i=e,o}\frac{ 1-\dfrac{2i}{\kappa a}\displaystyle\sum_{n=-\infty}^{\infty}F_p(k_i,\omega)}{  1+\dfrac{2i}{\kappa a}\displaystyle\sum_{n=-\infty}^{\infty}F_p(k_i,\omega)}\\
\label{rsnla}
r_s=&\frac{1}{2}\displaystyle\sum_{i=e,o}\frac{ 1+\dfrac{2i\kappa c^2}{a\omega^2}\displaystyle\sum_{n=-\infty}^{\infty}F_s(k_i,\omega)}{ -1+\dfrac{2i\kappa c^2}{a\omega^2}\displaystyle\sum_{n=-\infty}^{\infty}F_s(k_i,\omega)}\\
F_p(k_i,\omega)&\equiv\dfrac{1}{k_i^{2}}\left(\dfrac{q_i^2}{\epsilon_t(k_i,\omega)-c^2k_i^2/\omega^2}+\dfrac{p^2}{\epsilon_l(k_i,\omega)^2}\right)\\
F_s(k_i,\omega)&\equiv\dfrac{1}{\epsilon_t(k_i,\omega)-c^2k_i^2/\omega^2}\\
F_p(k,\omega)&\equiv\dfrac{1}{k^{2}}\left(\dfrac{q^2}{\epsilon_t(k,\omega)-c^2k^2/\omega^2}+\dfrac{p^2}{\epsilon_l(k,\omega)^2}\right)\\
F_s(k,\omega)&\equiv\dfrac{1}{\epsilon_t(k,\omega)-c^2k^2/\omega^2}
\end{align}
where $\kappa^2=\omega^2/c^2-p^2$, $q_e=2n\pi/a$, $q_o=(2n+1)\pi/a$, $k_e^2=p^2+q_e^2$, $k_o^2=p^2+q_o^2$, $p$ is the component of the photon wavevector in the plane of the half space, and $\epsilon_l(k,\omega)$ and $\epsilon_t(k,\omega)$ are the longitudinal and transverse components, respectively, of the Fourier decomposition of the nonlocal dielectric response. While the expressions (\ref{rpnla}) and (\ref{rsnla}) are valid for a general nonlocal dielectric response, for all numerical results presented in this paper we use the Lindhard forms
\begin{subequations}
\begin{align}
\epsilon_{l}(k,\omega)&=1+\frac{3\omega_{p}^{2}}{k^2v_F^2}\frac{(\omega+i\nu)%
f_{l}((\omega+i\nu)/kv_F)}{\omega+i\nu f_{l}((\omega+i\nu)/kv_F)},\\
\epsilon_t(k,\omega)&=1-\frac{\omega_p^2}{\omega(\omega+i\nu)}f_t((\omega+i\nu)/kv_F),
\end{align}
\end{subequations}
\begin{subequations}
\begin{align}
f_{l}(x)&=1-\frac{x}{2}\ln(x+1)/(x-1),\\
f_{t}(x)&=\frac{3}{2}x^2-\frac{3}{4}x(x^2-1)\ln(x+1)/(x-1).
\end{align}
\end{subequations}
Here $\nu$ is the electron collision frequency, $\omega_{p}=(4\pi ne^{2}/m)^{1/2}$
is the plasma frequency, and $v_{F}$ is the Fermi velocity. Eqs. (\ref{rpnla}) and (\ref{rsnla}) are the primary mathematical result of the present work, and are applied in the following sections to a variety of physical systems. This derivation runs closely parallel to that of Jones et al. \cite{joneskliewerfuchs} and Esquivel-Sirvent et al. \cite{esquivel}.

\section{Applications}
\label{app}
\subsection{Decoherenece}
\subsubsection{Energy relaxation}
\label{relax}
Here we present a quantitative comparison of the relaxation times of a point charge or spin qubit when exposed to a conducting half space or thin film in both a local and nonlocal treatment.  The relaxation rate for a charge or spin qubit is proportional to the spectral density of the fluctuating electric or magnetic field, respectively, at the location of the qubit.  Quantitatively, for a charge qubit of electric dipole moment $d$ or a spin qubit of magnetic dipole moment $\mu$ pointing in the $i$th direction at position $\textbf{r}$ with level separation $\omega_Z$, we have
\begin{align}
\label{T1}
\frac{1}{T_{1,c}}&=\frac{d^2}{\hbar^2} \chi^E_{ii}(\vec{r}, \vec{r}, \omega_Z) \coth\left(\frac{\hbar\omega_Z}{2k_{B}T}\right),  \\
\frac{1}{T_{1,s}}&=\frac{\mu^2}{\hbar^2} \chi^B_{ii}(\vec{r}, \vec{r}, \omega_Z) \coth\left(\frac{\hbar\omega_Z}{2k_{B}T}\right) ,
\end{align}
where $ \chi^{E,B}_{ii}(\vec{r}, \vec{r}, \omega_Z)$ are the electric and magnetic spectral densities, respectively, and $\vec{r}$ is the location of the qubit. The spectral densities are given by an integral expression involving the reflection coefficients.  If we take the the qubit to point along the $z$-direction, perpendicular to the surface, the relevant components of the spectral density tensors are
\begin{align}
\chi^E_{zz}(z, z, \omega) &=\hbar \text{Re}\int_{0}^{\infty}\frac{p^3}{q}dp e^{2iqz}r_{p}(p)\\
\chi^B_{zz}(z, z, \omega) &=\frac{\hbar}{c^2} \text{Re}\int_{0}^{\infty}\frac{p^3}{q}dp e^{2iqz}r_{s}(p)
\end{align}
where $q=\sqrt{\omega^2/c^2-p^2}$ for $p^2\leq\omega^2/c^2$ and $q=i\sqrt{p^2-\omega^2/c^2}$ for
$p^2>\omega^2/c^2$ is the $z$-component of the photon wavevector, and $p$ is the transverse component.

\begin{figure} 
\includegraphics[width = 1.0 \columnwidth] {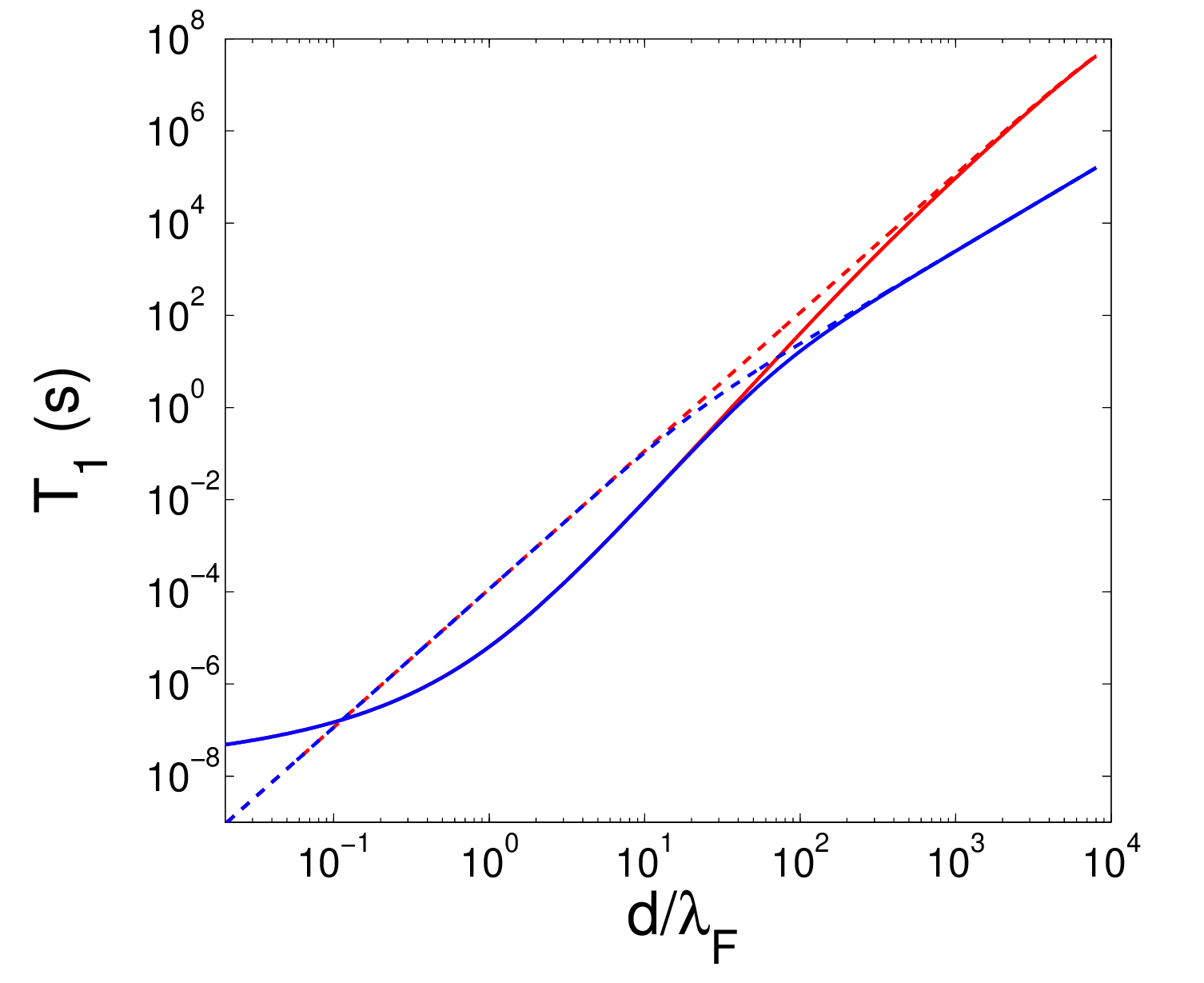}
\caption {Plot of $T_1$ time of a point charge qubit as a function of distance from the conductor, expressed in units of the Fermi wavelength, $\lambda_F$.  Dashed lines indicate a local and solid lines a nonlocal dielectric response.  Red curves are for a conducting halfspace and blue curves are for a thin film of thickness $a=10nm$.}
\label{T1vsz c} 
\end{figure}

\begin{figure} 
\includegraphics[width = 1.0 \columnwidth] {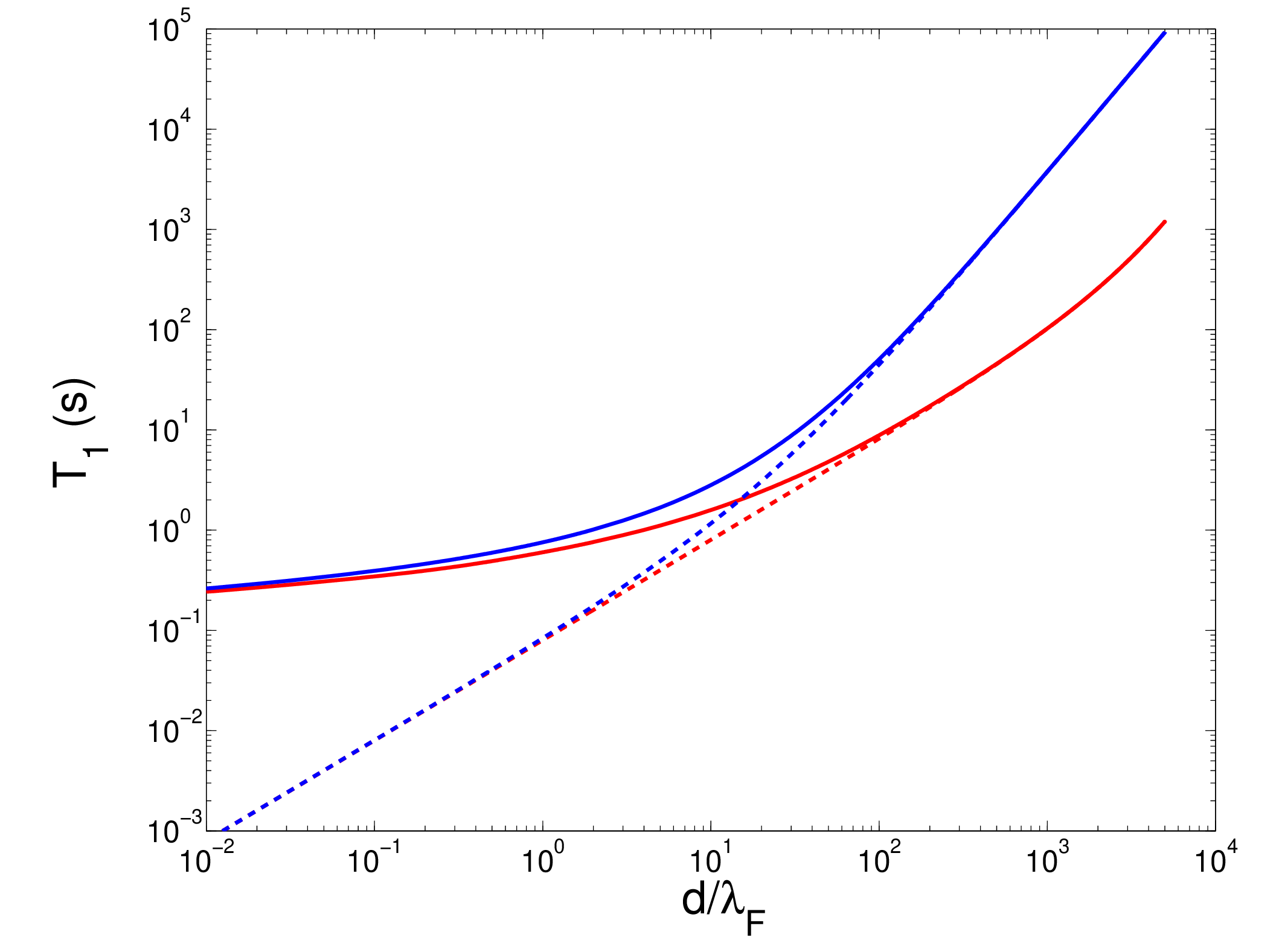}
\caption {Plot of $T_1$ time of a point spin qubit as a function of distance from the conductor, expressed in units of the Fermi wavelength, $\lambda_F$.  Dashed lines indicate a local and solid lines a nonlocal dielectric response.  Red curves are for a conducting halfspace and blue curves are for a thin film of thickness $a=10nm$.}
\label{T1vsz s} 
\end{figure}

Figure \ref{T1vsz c} shows the $T_1$ time for a charge qubit as a function of distance from the conductor.  Our primary result, $T_1$ from a thin film with a nonlocal dielectric function, is given by the solid blue curve.  Of particular note are its convergence to the nonlocal halfspace result as $z\rightarrow 0$ and its convergence to the local thin film result for large $z$.   For intermediate distances the nonlocal field fluctuations are enhanced above those given by a local treatment, while for smaller distances they converge to a finite value.  Also, for separations larger than the Fermi wavelength the electric field fluctuations outside a thin film are enhanced relative to those outside a halfspace. Figure \ref{T1vsz s} shows comparable results for a spin qubit, which will relax from fluctuations of the magnetic field. The enhancements of the nonlocal over the local field strength, and of the thin film over the halfspace, are not present for the magnetic case.

\subsubsection{Dephasing}
In this section we present results for the pure dephasing time of a charge or spin qubit from EWJN near a thin film.  The dephasing rate may be found through an examination of the off-diagonal elements of the time-dependent density matrix.  We assume initially that dephasing dominates over energy relaxation, and so consider qubit-environment coupling which is diagonal in the energy eigenbasis of the qubit.  Following \cite{mozyrsky}, our Hamiltonian takes the form
\begin{align}
H&=H_s+H_b+H_i\nonumber \\
&=\frac{1}{2}\sigma_z\omega_z+\sum_k\omega_k a_k^\dagger a_k+\Lambda_s\sum_k\left(g_k^*a_k+g_ka_k^\dagger\right),	
\end{align}
where $H_s$ is the two-level system Hamiltonian of the qubit, $H_b$ is the bath Hamiltonian for the fluctuating field, and $H_i$ represents the system-bath interaction.  $a_k^\dagger$ and $a_k$ are creation and annihilation operators, respectively, for field modes with wavevector $k$. $\Lambda_s$ is the coupling strength of the system observable to the fluctuating environment, and $g_k$ is the coupled field quantity with mode $k$.  In our case, $\Lambda_s$ will always be proportional to $\sigma_z$ in the pseudospin eigenbasis of the qubit. Because $H_i$ then commutes with $H_s$, our model will not describe energy relaxation. For the case of a charge qubit, the creation and annihilation operators are for electric field modes, while for a spin qubit, we have magnetic field creation and annihilation operators. For a charge qubit $\Lambda_s=d\sigma_z$, while for a spin qubit $\Lambda_s=\mu\sigma_z$, where $d$ and $\mu$ are the electric and magnetic dipole moments, respectively. If we take the qubit to point in the $i^{th}$ direction, $g_k=E_{k,i}$ for a charge qubit, where $E_{k,i}$ is the $i^{th}$ component of an electric field fluctuation with wavevector $k$, while $g_k=B_{k,i}$ for a spin qubit, where $B_{k,i}$ is the corresponding component of the magnetic field.  

As shown in \cite{mozyrsky}, the time dependence of the off-diagonal components of the reduced density matrix can be written as
\be
\rho_{01}(t)=\rho_{01}(0)e^{-\Gamma(t)},
\ee
where for a two-level system
\be
\label{dephaserate}
\Gamma(t)=\frac{1}{2\hbar^2}\sum_k\frac{|g_k|^2}{\omega_k^2}\sin^2\frac{\omega_kt}{2}\coth\frac{\beta\omega_k}{2}
\ee
The density matrix has been reduced in the sense of taking a thermal and quantum average over the bath degrees of freedom.  This allows $|g_k|^2$ to be expressed in terms of the electric and magnetic spectral densities, defined in Section \ref{relax}. The dephasing time $T_\phi$ is then defined as the value of $t$ for which $\Gamma(t)=1$.

A realistic qubit system will experience both pure dephasing and energy relaxation.  In this case the system-bath interaction Hamiltonian will contain terms proportional to $\sigma_x$ as well as $\sigma_z$ in the pseudospin basis.  It is a well-known result \cite{makhlin} that the dephasing time $T_2$ is then given by a reciprocal sum of contributions from energy relaxation and pure dephasing:
\be
\frac{1}{T_2}=\frac{1}{2T_1}+\frac{1}{T_\phi},
\ee
where $T_1$ is given in Section \ref{relax}.

\subsection{Stress-Energy Tensor}
In this section we present results for two closely related phenomena that are proportional to the stress-energy tensor in the vicinity of the conducting film.
\subsubsection{Heat Transfer}
In this section we calculate the heat transfer between thin films with a nonlocal response.  Heat flux from one surface to another will be proportional to the value of the Poynting vector in the direction of their separation at the location of the second surface.  We will thus be interested in the ensemble average of
\be
\langle \textbf{S}(\textbf{r})\rangle_\omega=\frac{c}{8\pi}\left(\langle \textbf{E}(\textbf{r})\times\textbf{B}^*(\textbf{r})\rangle_\omega+\langle \textbf{E}^*(\textbf{r})\times\textbf{B}(\textbf{r})\rangle_\omega\right)
\ee
Volokitin \cite{volokitin} found, for the case of two parallel semiinfinite bodies $1$ and $2$ with separation $z$ and reflection coefficients $r_{s1}, r_{p1}, r_{s2},$ and  $r_{p2}$,
\begin{align}
\label{sz}
S_z=&\hbar\omega\int_0^\infty\frac{d\omega}{2\pi}\left(N_1(\omega)-N_2(\omega)\right)\nonumber\\
&\times\Bigg(4\int_{q>\omega/c}\frac{d^2q}{(2\pi)^2}e^{-2|k|z}\nonumber\\
&\times\frac{\textrm{Im}r_{p1}(\textbf{q},\omega)\textrm{Im}r_{p2}(\textbf{q},\omega)}{\left(1-e^{-2|k|z}r_{p1}(\textbf{q},\omega)r_{p2}(\textbf{q},\omega)\right)^2}+[p\rightarrow s]\Bigg)
\end{align}
$[p\rightarrow s]$ denotes replacing the coefficients $r_p$ with $r_s$, and $N_{1,2}$ represent the Planck functions for the left or right film, respectively
\be
N_i(\omega)=\left(e^{\hbar\omega/k_BT_i}-1\right)^{-1},
\ee
In Eq. (\ref{sz}) we have dropped the lower portion of the integral over $q$ when $q<\omega/c$.  This part of the spectrum represents the radiative blackbody contribution to heat transfer, and by assumption it is negligibly small compared to the evanescent-wave contribution. 
If Eqs. (\ref{rpt}) and (\ref{rst}) are plugged into Eq. (\ref{sz}), a $1/z^2$ divergence in the heat transfer rate will emerge.  The nonlocal reflection coefficients for a thin film, Eqs. (\ref{rpnla}) and (\ref{rsnla}), vanish for sufficiently large wavevectors $q>1/\lambda_F$, which will remove this divergence to give a finite heating rate at zero separation.

\subsubsection{Casimir effect}
It is instructive to see how the inclusion of nonlocal dielectric properties affects the Casimir attraction between two parallel thin metallic films.  The Casimir interaction between thin films has been studied previously by several authors.  Beyond the treatment of Esquivel-Sirvent et al., mentioned in the introduction, Bostr$\ddot{\textrm{o}}$m et al. calculated the attractive force between atomically thin gold films, using density functional theory to derive the anisotropic deviations of the dielectric function of a thin film from its value in a bulk conductor \cite{bostrom}.  They found that the more accurate anisotropic dielectric function gives an enhanced attractive force relative to what is obtained using the bulk dielectric function.  The force, however, is still suppressed compared to the force between gold half-spaces. Their treatment of the dielectric function was local, however, and led to the usual unphysical divergence of the Casimir force at vanishing separation.  To calculate the Casimir force per area between thin films with a nonlocal response, we use the generalization of the Lifshitz formula derived by Moch$\acute{\textrm{a}}$n et al \cite{mochan}:
\begin{align}
\label{CasForce}
\frac{F(L)}{A}&=\frac{\hbar c}{2\pi^2}\int_0^\infty dQQ\int_{q\geq 0}dk\frac{\tilde{k}^2}{q}\nonumber\\
&\times \textrm{Re}\left(\frac{r_{s1}r_{s2}e^{2i\tilde{k}L}}{1-r_{s1}r_{s2}e^{2i\tilde{k}L}}+\frac{r_{p1}r_{p2}e^{2i\tilde{k}L}}{1-r_{p1}r_{p2}e^{2i\tilde{k}L}}\right)
\end{align}
Here $\tilde{k}=k+i0^+$, and the integral over $k$ runs from $iQ$ to $0$ and then to $\infty$.  The subscripts $1$ and $2$ on the reflection coefficients refer to the left and right surfaces, respectively. If the Fresnel reflection coefficients are plugged into Eq. (\ref{CasForce}), the original Lifshitz formula is recovered, but this expression is more generally applicable.

\begin{figure} 
\includegraphics[width = 1.0 \columnwidth] {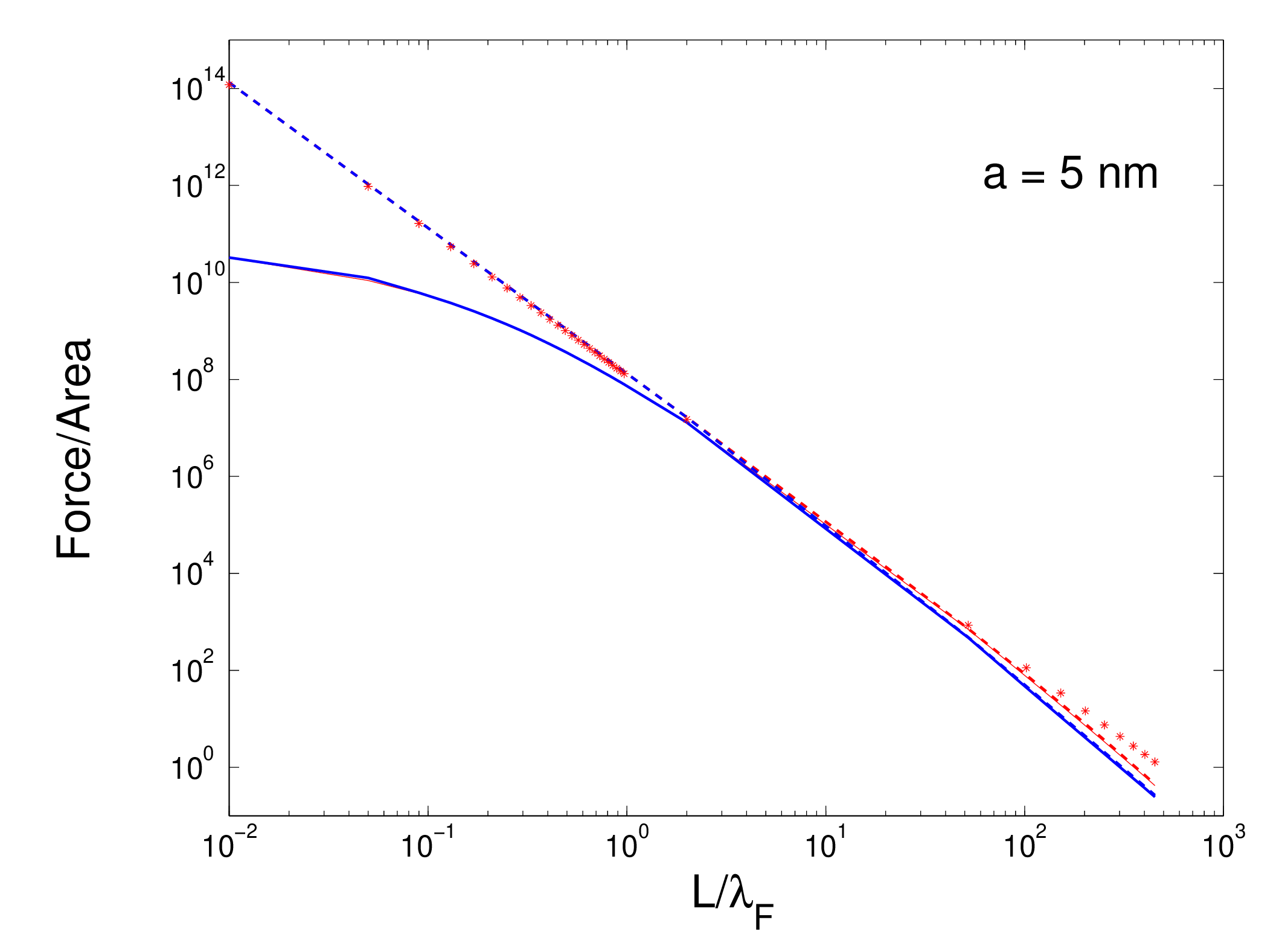}
\caption {Plot of the Casimir force per area between two metallic plates as a function of their separation, in units of the Fermi wavelength.  Dashed lines indicate a local and solid lines a nonlocal dielectric response.  Red curves are for two conducting halfspaces and blue curves are for two thin film of thickness $a=5$nm.}
\label{cas} 
\end{figure}

The Casimir forces per area between two thin films and two half-spaces in both a local and nonlocal treatment are shown in Figure \ref{cas}.  A nonlocal treatment of the dielectric function is shown to naturally yield a finite force at zero separation for both thin films and halfspaces, without having to use renormalization techniques or an external cutoff on high frequency modes.  The distinction between the Casimir attraction of thin films and halfspaces is insignificant at separations that are sufficiently small to necessitate a nonlocal dielectric function.  For larger separations we find the attraction between thin films is suppressed compared to the attraction between halfspaces, consistent with previous work \cite{bostrom}.

\section{Limiting cases}
\label{limit}
Including a nonlocal dielectric response alters the expressions for EWJN strength and the Casimir force soley through a modification of the reflection coefficients $r_p$ and $r_s$.  

Taking the limit $a\rightarrow\infty$ in  Eqs. (\ref{rpnla}) and (\ref{rsnla}) eliminates the distinction between the even and odd summations by converting them both into an integral over a continuous $q$.  Doing so gives the reflection coefficients for a metallic halfspace with a nonlocal dielectric response
\begin{align}
\label{rpnl}
r_{p}&=\frac{ 1-\dfrac{2i}{\pi \kappa}\displaystyle\int_0^\infty \frac{dq}{k^2}\left(\frac{q^2}{\epsilon_t(k,\omega)-c^2k^2/\omega^2}+\dfrac{p^2}{\epsilon_l(k,\omega)^2}\right)}{ 1+\dfrac{2i}{\pi \kappa}\displaystyle\int_0^\infty \frac{dq}{k^2}\left(\frac{q^2}{\epsilon_t(k,\omega)-c^2k^2/\omega^2}+\frac{p^2}{\epsilon_l(k,\omega)^2}\right)},\\
\label{rsnl}
r_s&=\frac{\dfrac{2i\kappa c^2}{\pi\omega^2}\displaystyle\int_0^\infty\frac{dq}{\epsilon_t(k,\omega)-c^2k^2/\omega^2}+1}{\dfrac{2i\kappa c^2}{\pi\omega^2}\displaystyle\int_0^\infty\frac{dq}{\epsilon_t(k,\omega)-c^2k^2/\omega^2}-1},
\end{align}
where $k^2=p^2+q^2$. Alternatively, the reflection coefficients for a thin film with a local response can be obtained from Eqs. (\ref{rpnla}) and (\ref{rsnla}) by setting $\epsilon_t(k,\omega)=\epsilon_l(k,\omega)=\epsilon(\omega)$, where $\epsilon(\omega)$ is a local dielectric function that is independent of $k$.  In this case the summations over $n$ may be evaluated in closed form and give
\begin{align}
r_p&=\frac{1}{2}\left(\frac{\kappa\epsilon-i\kappa_1\cot(\kappa_1a/2)}{\kappa\epsilon+i\kappa_1\cot(\kappa_1a/2)}+\frac{\kappa\epsilon+i\kappa_1\tan(\kappa_1a/2)}{\kappa\epsilon-i\kappa_1\tan(\kappa_1a/2)}\right)\nonumber \\
r_s&=\frac{1}{2}\left(\frac{\kappa-i\kappa_1\cot(\kappa_1a/2)}{\kappa+i\kappa_1\cot(\kappa_1a/2)}+\frac{\kappa+i\kappa_1\tan(\kappa_1a/2)}{\kappa-i\kappa_1\tan(\kappa_1a/2)}\right)\nonumber
\end{align}
where $\kappa_1^2=\omega^2\epsilon/c^2-p^2$. Combining the two terms yields the form given in \cite{poudel}:
\begin{align}
\label{rpt}
r_p &= \frac{\epsilon^2\kappa^2-\kappa_1^2 }{\kappa_1^2+\epsilon^2\kappa^2 +2i\kappa\kappa_1\epsilon \cot(\kappa_1a)} \\
\label{rst}
r_s &= \frac{\kappa^2-\kappa_1^2}{\kappa^2+\kappa_1^2+2i\kappa\kappa_1\cot(\kappa_1a)} 
\end{align} 
Finally, setting $\epsilon_t(k,\omega)=\epsilon_l(k,\omega)=\epsilon(\omega)$ in Eqs. (\ref{rpnl}) and (\ref{rsnl}) or sending $a\rightarrow\infty$ in Eqs. (\ref{rpt}) and (\ref{rst}) gives the traditional Fresnel reflection coefficients
\begin{align}
r_p = \frac{\epsilon \kappa-\kappa_1}{\epsilon \kappa+\kappa_1} \,, \quad \nonumber 
r_s = \frac{\kappa-\kappa_1}{\kappa+\kappa_1} \,. \nonumber 
\end{align}

\section{Conclusions}
\label{conc}
We have presented a detailed microscopic treatment of the reflective properties of thin metallic films, where the use of a general nonlocal dielectric function has incorporated the discrete nature of the valence electrons inside the metal.  The inclusion of nonlocality in the dielectric function removes a spurious divergence in the rate of heat transfer and the strengths of both evanescent-wave Johnson noise and the Casimir attraction at zero separation from the film.  This is accomplished through a suppression of the reflection coefficients for values of the transverse wavevector larger than the inverse of the interatomic separation. Uniquely, electric field evanescent-waves are enhanced in the nonlocal treatment for an intermediate range of distances on the order of the Fermi wavelength.  This enhancement will lead to a decrease in the decoherence times of charge qubits below what would be expected from a local treatment. This comes about from an enhancement of $r_p$ that is not present for $r_s$. Additionally, there is an enhancement of $r_p$ for a thin film over that for a halfspace for all separations.  This enhancement can be understood by analogy to a quantum particle trapped in a finite square well.  Further squeezing of the particle will lead to increased leakage of the wavefunction into the forbidden region.  Because the material is not magnetoactive, this enhancement is not present for magnetic field fluctuations, i.e., for $r_s$. 

We expect the nonlocal corrections to the reflection coefficients to become more practically relevant in the future as micromechanical devices are further miniaturized.  Beyond the use of metallic sheets in such devices, the unique electromagnetic properties of graphene have made it a popular material in the development of micromechanical devices \cite{messina} *more graphene references*.  It would be interesting to see how the results presented here would generalize to the case of graphene.  At present, decoherence times in contemporary qubit devices seem to be limited by EWJN only for spin qubits at low external magnetic field \cite{poudel}. However, in the future we expect EWJN to become a dominant source of decoherence in charge qubits also as other noise sources are suppressed.

This work was funded by ARO and LPS Grant No. W911NF-11-1-0030.


\begin{thebibliography}{10}

\bibitem{henkel} C. Henkel, S. P$\ddot{\text{o}}$tting, and M. Wilkens, Appl. Phys. B \textbf{69}, 379 (1999).

\bibitem{borselli} M. G. Borselli, K. Eng, E. T. Croke, B. M. Maune, B. Huang, R. S. Ross, A. A. Kiselev, P. W. Deelman, I. Alvarado-Rodriguez, A. E. Schmitz, M. Sokolich, K. S. Holabird, T. M. Hazard, M. F. Gyure, and A. T. Hunter, Appl. Phys. Lett., \textbf{99}, 063109 (2011).

\bibitem {fordweber} G. W. Ford and W.H. Weber, Phys. Rep. \textbf{113}, 195 (1984).

\bibitem {LLSP2} E. M. Lifshitz and L. P. Pitaevskii, Statistical Physics,
vol.IX, Part 2, (Pergamon, 1980) Chapter VIII.

\bibitem{LLECM} L. D. Landau and E. M. Lifshitz, Electrodynamics of Continuous Media, vol. XIII, (Pergamon, 1960) Chapter X.

\bibitem {joneskliewerfuchs} W. E. Jones, K. L. Kliewer, and R. Fuchs, Phys. Rev. \textbf{178}, 3 (1969).

\bibitem {langsjoen} L. S. Langsjoen, A. Poudel, M. G. Vavilov, and R. Joynt, Phys. Rev. A, \textbf{86}, 010301 (2012).

\bibitem {poudel} A. Poudel, L. S. Langsjoen, M. G. Vavilov, and R. Joynt, Phys. Rev. B, \textbf{87}, 045301 (2013).

\bibitem{volokitin} A.I. Volokitin and B. Persson, Rev. Mod. Phys. \textbf{79}, 1291 (2007).

\bibitem{mozyrsky}  D. Mozyrsky and V. Privman, J. of Stat. Phys., \textbf{91}, pp. 787-799 (1998).

\bibitem{palma} M. G. Palma, K. A. Suominen, and A. K. Ekert, Proc. R. Soc. London, Ser. A \textbf{452}, 567 (1996).

\bibitem{makhlin} Yu. Makhlin, G. Schn, and A. Shnirman, New Directions in Mesoscopic Physics (Towards Nanoscience), pp. 197-224 (2003).

\bibitem{bostrom} M. Bostr$\ddot{\textrm{o}}$m, C. Persson, and B. E. Sernelius, Eur. Phys. J. B, 86: 43, (2013).

\bibitem{esquivel} R. Esquivel-Sirvent and V. B. Svetovoy, Phys. Rev. B, \textbf{72}, 045443 (2005).

\bibitem{mochan} W. L. Moch$\acute{\textrm{a}}$n, Rev. Mex. Fis., \textbf{48} (4) pp. 339-342 (2002).

\bibitem{messina} R. Messina and P. Ben-Abdallah, Scien. Rep. \textbf{3}, 1383 (2013)

\end{thebibliography}
\end{document}